\documentclass[review]{elsarticle}

\journal{Journal of \LaTeX\ Templates}

\begin{document}

\begin{frontmatter}

\title{Near-infrared holographic photorefractive recording under applied electric field in undoped Bi$_{12}$TiO$_{20}$ sillenite crystal}



\author[1]{Andr\'e L. Moura}
\author[2]{Alexsandro F. Pereira}
\author[1,3]{Askery Canabarro}
\author[4]{Jesiel F. Carvalho}
\author[5]{Ivan de Oliveira}
\author[2]{Pedro V. dos Santos\corref{mycorrespondingauthor}}
\cortext[mycorrespondingauthor]{Corresponding author}
\ead{pedro@fis.ufal.br}

\address[1]{{Grupo de F\'isica da Mat\'eria Condensada, N\'ucleo de Ci\^encias Exatas, Campus Arapiraca, Universidade Federal de Alagoas, 57309-005, Arapiraca-AL, Brazil}}
\address[2]{Instituto de F\'isica, Universidade Federal de Alagoas, 57072-970, Macei\'o-AL, Brazil}
\address[3]{International Institute of Physics, Federal University of Rio Grande do Norte, 59070-405 Natal, Brazil}
\address[4]{Instituto de F\'isica Universidade Federal de Goi\'as, Goi\^ania, GO, Brazil}
\address[5]{GOMNI-Grupo de \'Optica e Modelagem Num\'erica, Faculdade de Tecnologia/UNICAMP, Limeira, SP, Brazil}

\begin{abstract}
Holographic recording in Bi$_{12}$TiO$_{20}$ crystal at 1064 nm is investigated aiming the characterization of diffraction efficiency under action of applied dc electric field ($E_0$). An enhancement of 12-fold in the diffraction efficiency was revealed when $E_0$ increased from 0.0 to 4.2 kV/cm. The theoretical dependence of the diffraction efficiency upon $E_0$ was investigated using the standard model for photorefractivity and the results showed a good experimental data fitting, allowing the computation of the acceptor concentration in the crystal of $\approx$ $5.3\times 10^{15}$ cm$^{-3}$.
\end{abstract}


\end{frontmatter}


\section{Introduction}

Photonics requires the characterization of optical properties of a myriad of materials, as glasses, crystals, and nanocrystal powders. Among the crystalline ones, the photorefractives are interesting for hologram recording due to the photoconductive and electro-optic effects which allow registering a pattern of light as a modulation of the refractive index \cite{gunter1989}. Sillenites (Bi$_{12}$MO$_{20}$ with M = Si, Ge or Ti) have been extensively characterized in the visible spectral range. Regarding longer wavelengths, the Bi$_{12}$TiO$_{20}$ (BTO) crystal allows holographic recording in the near infrared region of the electromagnetic spectrum where both optical activity and optical absorption are lower than in the visible region \cite{Sturman94,dosSantos2005}, and presents high effective electro-optic coefficient $(\approx 5.5$ pm/V) \cite{MOURA2013197}.


In \cite{Sturman94} and \cite{dosSantos2005} the authors investigated stationary (nonmoving) photorefractive hologram recorded in undoped BTO crystal at 1064 nm and 780 nm, respectively. These two experiments were carried out with and without a mechanism of pre-exposure to light, but no external electric field was applied to the samples, and in both experiments the results revealed recorded holographic grating very weak in spite of the diffraction efficiency measured with pre-exposure was approximately 3 times greater than that without pre-exposure. Another way to improve the recording characteristics at 1064 nm or 780 nm is by applying an ac or dc electric field to the sample, but there are not experimental results in the literature that report this fact at those wavelengths to the best of our knowledge. In fact, it is well known that photorefractive holograms produced under the action of externally applied electric field ($E_0$) exhibit comparatively large diffraction efficiencies \cite{micheron76,tarev79}. However, the action of the $E_0$ perturbs the crystal index ellipsoid due to the electro-optic effect, inducing a birefringence in the medium making the diffracted beams elliptically polarized; this phenomena produces additional difficulties to separate the diffracted beams from the signal and pump ones and consequently making difficult to determine the diffraction efficiency and also to improve the signal-to-noise ratio (SNR) by simply doing the use of a polarization analyzer \cite{dosSantos2005,micheron76,Marra86}. In order to overcome this difficult some phase modulation techniques that are applied to one of the recording beams have been proposed \cite{HUIGNARD1981249,refregier85,STEPANOV1985292,donatti09}.
In this work we take advantage of phase-sensitive detection and measure diffraction efficiency by direct chopping of elliptically polarized diffracted beams in photorefractive hologram recorded in undoped BTO crystal at 1064 nm under the action of a dc $E_0$. The effects of $E_0$ on the diffraction efficiency were experimentally and theoretically investigated.

\section{Experiment}

\begin{figure}[htbp]
\centering
\fbox{\includegraphics[width=\linewidth]{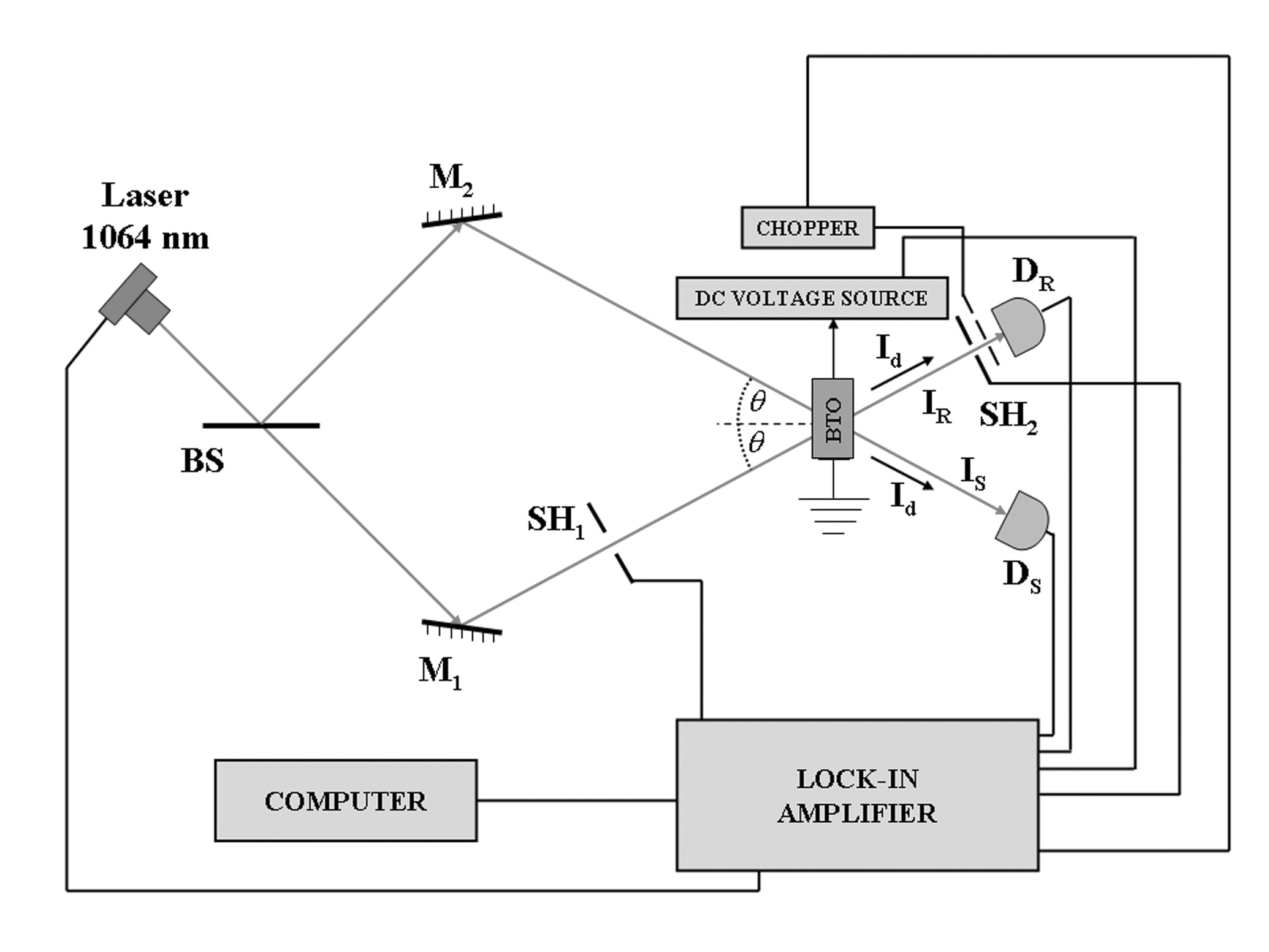}}
\caption{Holographic arrangement: BS is beam-splitter; M$_1$ and M$_2$ are mirrors; SH$_1$ and SH$_2$ are shutters; $I_R$, $I_S$ and $I_d$, are, respectively, the pump, signal and diffracted beams at the BTO crystal output; $D_R$ and $D_S$ are photodetectors.}
\label{fig:1}
\end{figure}

\begin{figure}[htbp]
\centering
\fbox{\includegraphics[width=\linewidth]{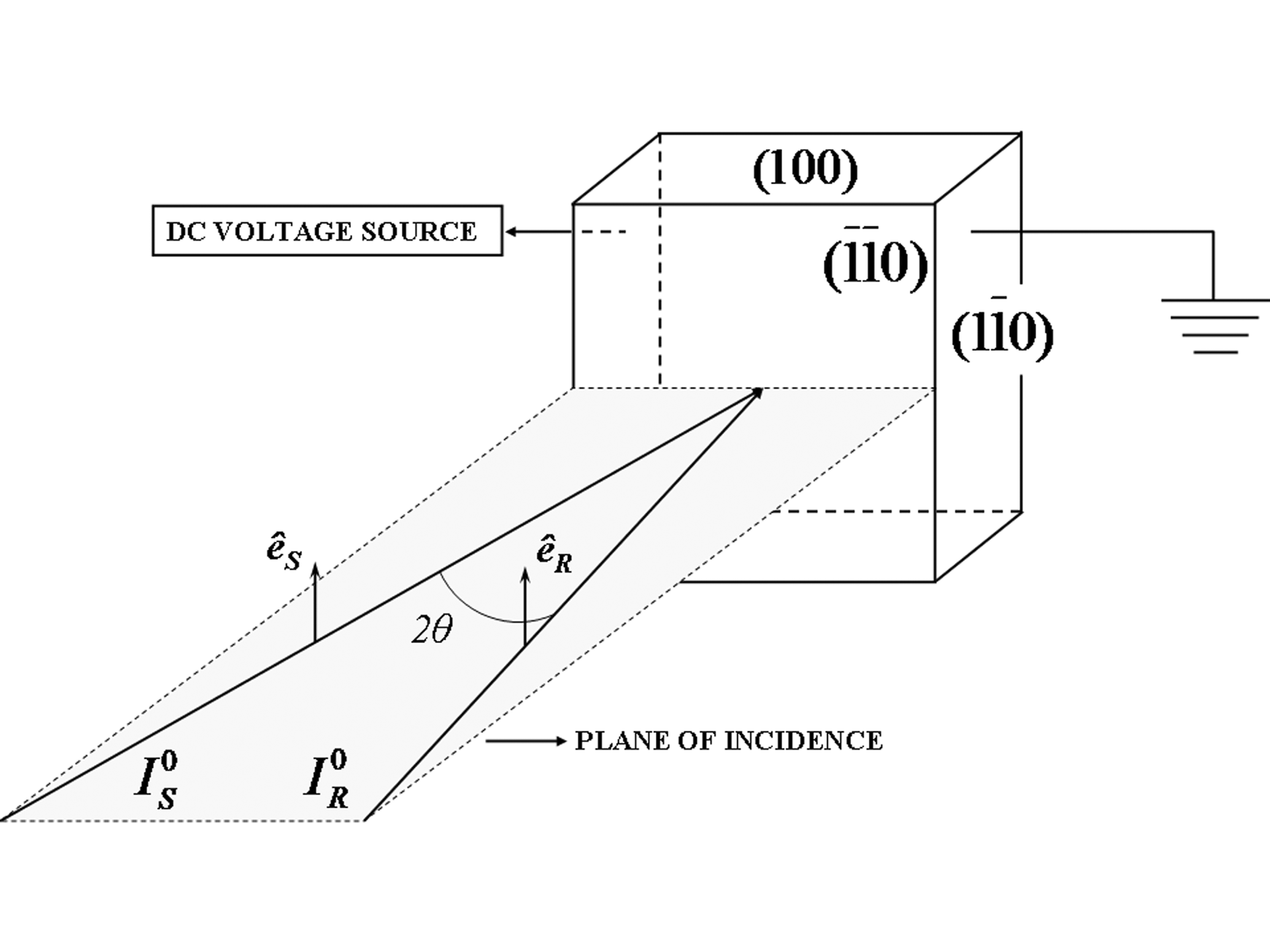}}
\caption{BTO crystal orientation. Intensities ($I_S^0$ and $I_R^0$) and polarizations ($\widehat{e}_S$ and $\widehat{e}_R$) of the recording beams at the input face of the sample.}
\label{fig:2}
\end{figure}

Holograms were recorded in an undoped BTO sample using $\lambda =1064$ nm beams from a 50 mW solid state laser in the interferometric arrangement shown in Fig. \ref{fig:1}. The intensities of the recording beams measured on the input face of the sample corresponded to a total intensity of 0.45 W/cm$^2$, with the signal beam intensity being 1/36 from the pump beam intensity. The diameter of the unexpanded beam is 2 mm. The BTO sample used measures 3.6 mm × 8.4 mm × 8.7 mm along the $[\overline{1} \hspace{0.05cm}\overline{1} \hspace{0.05cm} 0]$, the $[1 \hspace{0.05cm}\overline{1} \hspace{0.05cm} 0]$, and the $[0 \hspace{0.05cm}0\hspace{0.05cm} 1]$ directions, respectively. The light interference pattern produced by the two beams (with the electrical field vector being perpendicular to the plane of incidence) was always projected onto the $(\overline{1} \hspace{0.05cm}\overline{1} \hspace{0.05cm} 0)$-crystal face as depicted in Fig. \ref{fig:2}, with the $[0 \hspace{0.05cm}0\hspace{0.05cm} 1]$-crystal axis being perpendicular to the incidence plane and to the recorded grating vector \textbf{K}, with $|\textbf{K}| = 2\pi/\Lambda$, where $\Lambda = (\lambda/2) \sin\theta$ is the spatial frequency of grating, $\lambda$ is the laser wavelength, and $2\theta$ is the angle between the recording beams. In order to ensure similar initial conditions for the sample, prior to each experimental run it was always illuminated for 25 minutes with one ($I^0_S$: signal beam intensity) of the two 1064 nm recording beams while the other one ($I^0_R$: pump beam intensity) remained blocked by the shutter SH$_1$. In this experiment the shutters were adjusted to run in opposite modes, i.e., when SH$_1$ is closed (opened) SH$_2$ is simultaneously opened (closed). After that time of 25 minutes, SH$_1$ is opened and the light interference pattern writes a photorefractive hologram in the sample producing diffracted beams ($I_d$) in the $I_S$ and $I_R$ directions. The intensity $I_S$ ($I_R$) plus the diffracted signal in its direction are measured by the photodetector $D_S$ ($D_R$). When the shutter SH$_1$ is closed again (SH$_2$ is opened) the sample is illuminated uniformly by the signal beam,  causing the erasure of the hologram and consequently producing an in-Bragg diffracted beam which is chopped at 1 kHz and measured by the photodetector $D_R$. The photocurrents generated in $D_S$ and $D_R$ are fed into the lock-in amplifier connected to a computer. Calibration curves of the light intensities measured before $D_S$ and $D_R$ versus the corresponding lock-in voltages were prepared in order to directly compute the absolute value of the diffraction efficiency as $\eta = I_d/(I_d + I_S)$ and measured at the beginning of erasure (bulk absorption and interface losses are neglected). An external dc voltage is applied between the silver electrodes painted on the side surfaces of the sample, $(1 \hspace{0.05cm}\overline{1} \hspace{0.05cm} 0)$ and $(\overline{1} \hspace{0.05cm} 1 \hspace{0.05cm} 0)$ faces, making the resultant external electric field always perpendicular to the $[0 \hspace{0.05cm}0\hspace{0.05cm} 1]$-crystal axis. The spatial frequency of the recorded grating is $\Lambda = 1.55$ $\mu$m. As can be seen the setup presented here is identical to those normally used to record holograms in a two-wave mixing (TWM) configuration. The difference here is the lock-in amplifier used to detect the chopped weak diffracted signal behind the sample, and also to control others parts of the experiment as the shutters and the dc voltage source.

\section{Results and discussion}

\begin{figure}[htbp]
\centering
\fbox{\includegraphics[width=\linewidth]{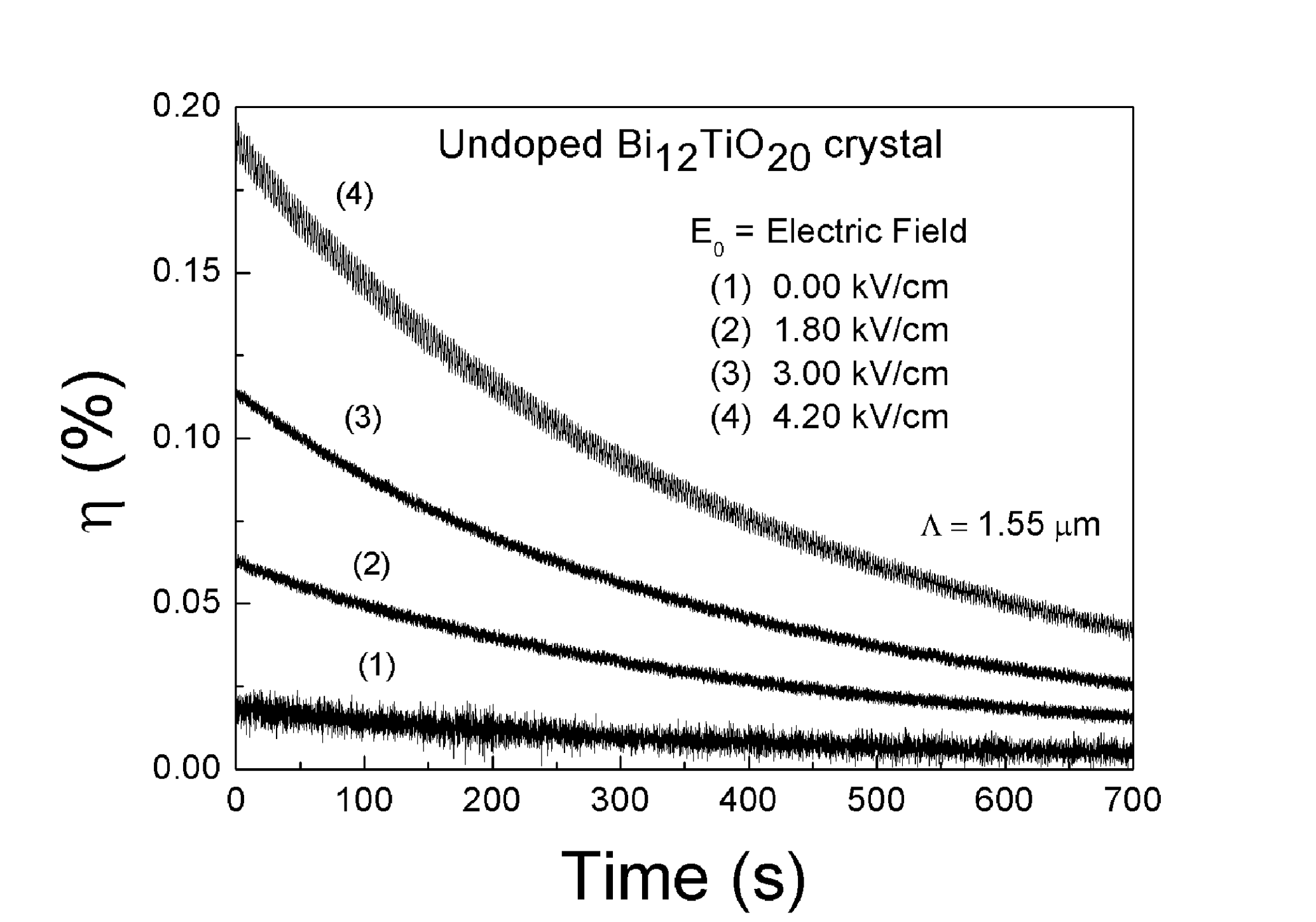}}
\caption{Diffraction efficiency as a function of time during the hologram erasure for different applied electric field.}
\label{fig:3}
\end{figure}

\begin{figure}[htbp]
\centering
\fbox{\includegraphics[width=\linewidth]{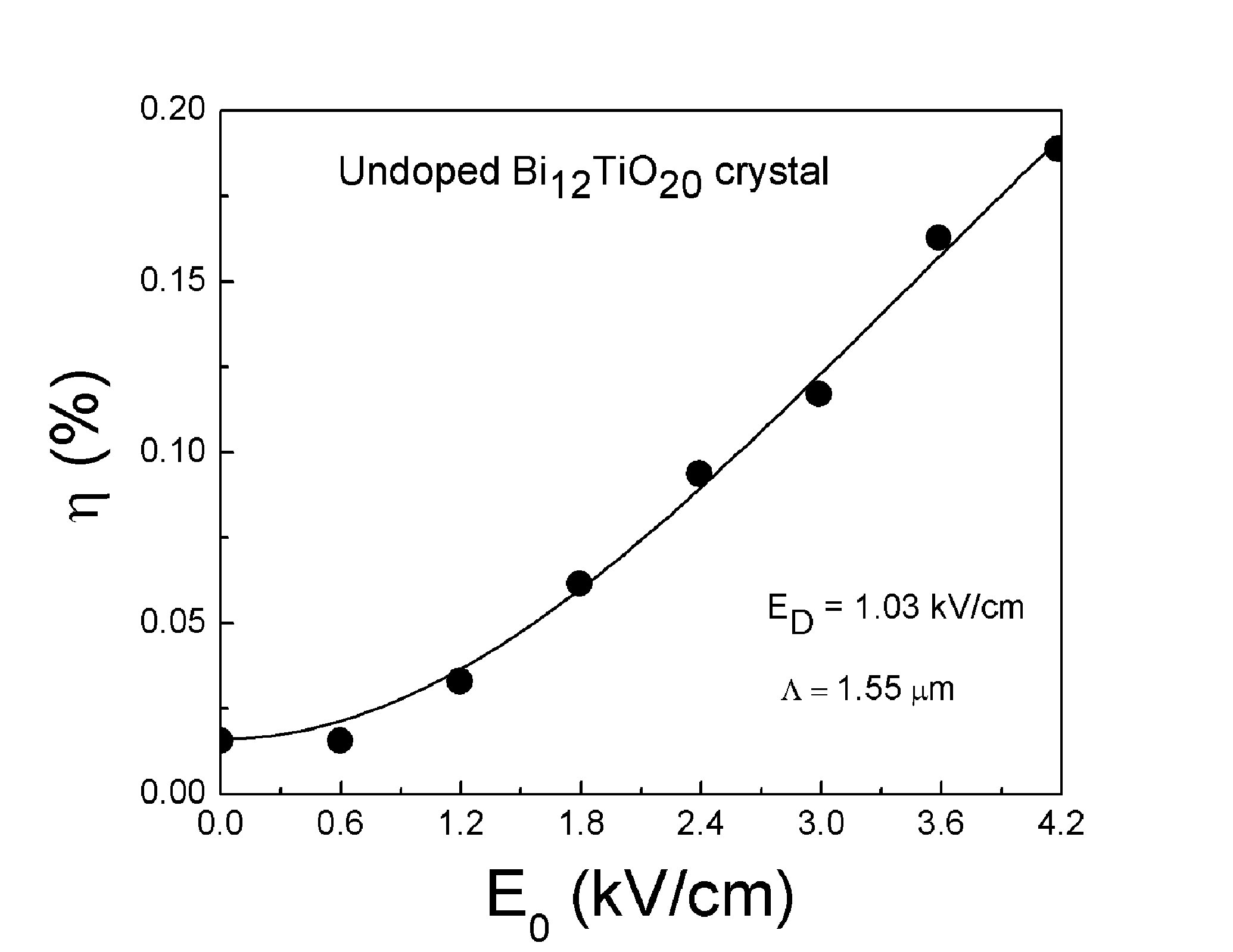}}
\caption{Diffraction efficiency as a function of the applied electric field. The solid line represents the theoretical fit obtained from Eq. 4 adjusted to the experimental data (symbols).}
\label{fig:4}
\end{figure}

After holographic recording for 5 minutes using the setup shown in Fig. \ref{fig:1}, the reference beam is blocked by SH$_1$ and the signal one is used at the same time to uniformly erase the hologram and as an in- Bragg probe beam to measure the time evolution of diffraction efficiency $\eta$. The time evolution of the chopped diffracted beam intensity was measured at the BTO sample output for different amplitudes of $E_0$ and is shown in Fig. \ref{fig:3}. Clearly is observed an enhancement in the diffracted signal when the amplitude of $E_0$ is increased. This enhancement is attributed to the increase of the photorefractive response of the BTO crystal to $E_0$ \cite{micheron76,tarev79}. It is important to point out here that measurements of diffraction efficiencies were also carried out for different chopping frequencies of 200 Hz, and 2 kHz, and no modification in the time evolution of the diffraction efficiency was observed (results not shown here). An analysis of the time decay curves depicted in Fig. \ref{fig:3} showed that they can be well fitted with a single exponential function indicating the probable participation of an unique active center \cite{frejlich2007} responsible for holographic erasure. The dependence of the diffraction efficiency upon $E_0$ was examined and the result is depicted (symbols) in Fig. \ref{fig:4}. As can inferred from the experimental data, the diffraction efficiency increased by a factor of 12-fold when the amplitude of the $E_0$ increased from 0 to 4.2 kV/cm. 
The enhancement of the diffraction efficiency as a function of $E_0$ was theoretically investigated by using the standard model \cite{gunter2005} and taking into account that only one photoactive center is participating in the photorefractive recording and erasure processes. A recorded photorefractive hologram (or diffraction grating) may be read by a beam which is Bragg-matched to the grating and this beam is scattered with diffraction efficiency (for $\eta \ll 1$) \cite{frejlich2007}

\begin{equation}
\eta \propto |E_{sc}|^{2},
\label{eq:1}
\end{equation}
where $E_{sc}$ is the space-charge-field \cite{gunter2005}

\begin{equation}
E_{sc} = \imath m \frac{E_D+ \imath E_0}{1 + \frac{E_D+ \imath E_0}{E_q} },
\label{eq:2}
\end{equation}
with $m$ being the visibility of the fringes of interference, $E_D=2 \pi k_B T/e\Lambda$, and $E_q=(e\Lambda N_A^{(0)})/(2\pi \varepsilon \varepsilon_0)$, where $k_B$ is the Boltzmann constant, $T$ the absolute temperature, $e$ the elementary charge, $N_A^{(0)}$ the acceptor concentration in the absence of illumination, $ε$ the dielectric constant, and $\epsilon_0$ is the dielectric permittivity of vacuum. It is straightforward to obtain that

\begin{equation}
\eta \propto m^ 2 \frac{E_0^2 + E_D^ 2}{(E_0/E_q)^2 + [(E_q+ E_D)/E_q]^2}.
\label{eq:3}
\end{equation}

In this approach the accumulative effect of charges in the shadow regions near the side surfaces electrodes \cite{klein88,grunnet95} was not take in account. Such effect contributes to decrease (screen) the action of the $E_0$ in the bulk of the material \cite{klein88,grunnet95}. In order to take into account the effect of screening produced by that charge distribution, we introduced a parameter $\xi$, called field factor, which can vary from 0 (no applied electric field) to 1 (no screening effect) and the Eq. \ref{eq:3} can now be written as

\begin{equation}
\eta \propto m^ 2 \frac{(\xi E_0)^2 + E_D^ 2}{(\xi E_0/E_q)^2 + [(E_q+ E_D)/E_q]^2}.
\label{eq:4}
\end{equation}

Equation \ref{eq:4} was fitted to the experimental data and the result is indicated by the solid line in plot of Fig. \ref{fig:4}, showing a good agreement between the theoretical curve and experimental data. Using $E_D = 1.03$ kV/cm ($T = 295$ K) we find $\xi\simeq 1$ and $E_q \simeq 5.0$ kV/cm. Using this value of $E_q$ and $\varepsilon = 47$ \cite{kamshilin96} we computed an acceptor concentration in the crystal of $N_A^{(0)} \simeq 5.3 \times 10^{15}$ cm$^{-3}$. This value is of the same order of magnitude than that one obtained by Odoulov et al. at 1064 nm \cite{Sturman94}, but that was estimated by using the technique proposed by Klein and Valley \cite{klein96} which has an experimental difficulty that is the necessity to vary the angle $2\theta$ between the recording beams in order to indirectly compute the acceptor concentration. In our approach presented here that difficulty does not exist since the angle $2\theta$ remains fixed. The field factor value $\xi \approx 1$ indicates that the influence of the screening effect is very weak and it seems not to be a limiting factor of the enhancement of the diffraction efficiency as a function of $E_0$ in the recording region at 1064 nm.

\section{Conclusion}

We measured diffraction efficiency by direct chopping of elliptically polarized diffracted beams in photorefractive holographic grating recorded in undoped Bi$_{12}$TiO$_{20}$ crystal at 1064 nm under action of an applied dc electric field ($E_0$)using a two-wave mixing configuration. The experimental results showed a 12-fold enhancement in the diffraction efficiency when $E_0$ was ranged from 0 to 4.2 kV/cm. This enhancement was attributed to photorefractive response of the material to $E_0$ and was investigated theoretically using the standard model for photorefractivity and the results showed to agree quite well with experimental data.

\section*{Acknowledgements}
We acknowledge financial support from the Brazilian Agencies: Conselho Nacional de Desenvolvimento Cient\'{i}fico e Tecnol\'{o}gico (CNPq), and Funda\c{c}\~ao de Amparo \`{a} Pesquisa do Estado de Alagoas (FAPEAL). AC acknowledges the financial support from MEC/UFRN.




\section*{References}

\bibliography{mybibfile}
\bibliographystyle{plain}

\end{document}